\documentclass[%
twocolumn,
superscriptaddress,
preprintnumbers,
 amsmath,amssymb,
aps,
prd,
floatfix,
]{revtex4-2}

\usepackage{graphicx,color}
\usepackage{dcolumn}
\usepackage{bm}
\usepackage[normalem]{ulem}
\usepackage{color} 
\usepackage{amsmath}
\usepackage{ytableau}
\usepackage{here}
\newcommand{\red}[1]{\textcolor{red}{#1}} 
\definecolor{red}{rgb}{0,0,0}

\begin{document}


\title{Multiquark clustering in neutron-star matter from color-spin molecular dynamics}

\author{Nobutoshi Yasutake}
\email{nobutoshi.yasutake@p.chibakoudai.jp}
\affiliation{Department of Physics, Chiba Institute of Technology, 2-1-1 Shibazono, Narashino, Chiba 275-0023, Japan}
\affiliation{Advanced Science Research Center, Japan Atomic Energy Agency, Tokai, Ibaraki 319-1195, Japan}

\author{Yuta Mukobara}
\email{mukobara.y.aa@m.titech.ac.jp}
\affiliation{Department of Transdisciplinary Science and Engineering, School of Environment and Society, Institute of Science Tokyo, Meguro, Tokyo, Japan}

\author{Aaron Park}
\email{aaron.park@yonsei.ac.kr}
\affiliation{Department of Physics and Institute of Physics and Applied Physics, Yonsei University, Seoul 03722, Korea}

\author{Su Houng Lee}
\email{suhoung@yonsei.ac.kr}
\affiliation{Department of Physics and Institute of Physics and Applied Physics, Yonsei University, Seoul 03722, Korea}

\author{Toshiki Maruyama}
\email{maruyama.toshiki@jaea.go.jp}
\affiliation{Advanced Science Research Center, Japan Atomic Energy Agency, Tokai, Ibaraki 319-1195, Japan}


\date{\today}

\begin{abstract}
We study the equation of state of neutron-star matter with color-spin molecular dynamics. The calculation includes the internal color and spin degrees of freedom and their time evolution. The matter composition, including strangeness under $\beta$ equilibrium, is determined by energy minimization. We find two main trends. First, within the present color-spin molecular dynamics framework and under the adopted clustering criterion along the stable neutron-star branch, isolated quarklike configurations do not appear; instead, color-magnetic interactions favor the self-consistent formation of multiquark clusters. Within the same criterion, the cluster-size distribution is concentrated at quark numbers that are multiples of three, corresponding to integer baryon numbers. \red{Second, relative to the conventional no-$K^*$ baseline, the interaction between strange and light quarks has a strong impact on neutron-star radii. This suggests that future radius measurements, together with phenomenological information on the strangeness-onset density, may help constrain flavor-sector interactions involving strangeness.}
\end{abstract}

\pacs{
26.60.+c,  
24.10.Cn,  
97.60.Jd,  
12.39.-x  
}

\maketitle

\section{\label{sec:level1}Introduction}
Despite decades of work on the equation of state (EOS) of dense matter in neutron-star (NS) interiors, fundamental challenges---most notably the sign problem---still leave large uncertainties. Recent observational advances, such as the Neutron Star Interior Composition Explorer~(NICER)~\cite{Riley2019, Miller2019, Riley2021, Miller2021, Vinciguerra2024, Choudhury2024} and gravitational-wave detectors~\cite{Abbott2017}, have steadily tightened constraints on NS mass-radius relations. The central question, therefore, is which physical properties of the EOS can be robustly extracted from the current constraints.

The most famous observational constraint is that the maximum NS mass is at least $\sim 2 M_\odot$~\cite{Antoniadis2013}. This condition is particularly relevant to the so-called ``{\it hyperon puzzle}'': it is well known that once strangeness is included, the EOS often becomes too soft to support a $2\,M_\odot$ star. In this context,
many-body effects in the medium are frequently proposed to resolve this issue, e.g. augmenting realistic two-body forces with phenomenological three-body terms~\cite{Nishizaki2002,Vidana2011,Yamamoto2014,Lonardoni2015,Togashi2016}, and using mean-field frameworks with explicit density dependence~\cite{Bednarek2012,Weissenborn2012,Muto2022,Ma2023,Huang2025}. 
For comprehensive overviews of microscopic and phenomenological EOS, the hyperon puzzle, and three-body forces, see Burgio et al.~\citep{Burgio2021}. In particular, recent work shows that the hyperon puzzle persists even when the interactions are partially constrained by ab initio lattice-QCD results~\cite{Vidana2025}. 
From a different perspective, medium effects---currently inaccessible to lattice-QCD calculations---could be a key ingredient. In our earlier molecular-dynamics (MD) study, we also implemented analogous many-body physics by adding interactions with power-law dependence on the interparticle separation~\cite{Akimura2005,Yasutake2024}.

Constraints on NS radii have tightened steadily, but a definitive bound has not yet been established. 
For example, current NICER-only observations for PSR J0030+0451 using a single-temperature plus protruding-single-temperature hot-region model suggest $M= 1.37 \pm 0.17 M_\odot$ and $R = 13.11 \pm 1.30$ km, while joint analyses including XMM-Newton data yield two different solutions: $M= 1.40^{+0.13}_{-0.12} M_\odot$ with $R = 11.71^{+0.88}_{-0.83}$ km, and $M= 1.70^{+0.18}_{-0.19} M_\odot$ with $R = 14.44^{+0.88}_{-1.05}$ km~\cite{Vinciguerra2024}. 
For PSR J0437-4715, NICER infers $M= 1.418 \pm 0.037 M_\odot$ and $R= 11.36^{+0.95}_{-0.63}$ km~\cite{Choudhury2024}.
\red{
In many nucleonic EOS studies, the symmetry energy and its density dependence are often discussed as ingredients that can influence neutron-star radii, although the strength of this connection is model dependent~\cite{Vinas2014,Li2019,Hornick2018}. Meanwhile, Wanajo et al. argued from black-hole-neutron-star merger nucleosynthesis that the ejecta require $Y_e \gtrsim 0.05$ and thereby favor an EOS similar to the density-dependent relativistic mean-field parametrization DD2~\cite{Wanajo2024}; this may be interpreted as support for comparatively strong symmetry-energy effects in the relevant density range. Since DD2 yields a comparatively large canonical radius, $R=13.2$ km at $M=1.4\,M_\odot$~\cite{DD2}, such comparatively large radii remain viable within current observational bounds.}
Recent astrophysical constraints on the EOS are comprehensively reviewed by Chatziioannou et al.~\cite{Chatziioannou2025}.

This paper also addresses the \textit{hyperon puzzle} within a quark molecular-dynamics framework. However, it is highly nontrivial to map realistic baryon-baryon interactions to quark-quark interactions~\cite{Yamamoto2022}. 
In our earlier MD study, we modeled the physics via quark-quark interactions mapped to the mesonic channels of the $\sigma$-$\omega$-$\rho$-$\phi$ quark-meson-coupling model; the implementation accounted only for light-quark pairs $qq$ $(uu,ud,dd)$ and $ss$ pairs, inadvertently neglecting the strange-light $sq$ channel~\cite{Akimura2005}. 
In this paper, we extend the setup to a $\sigma$-$\omega$-$\rho$-$\phi$-$K^\ast$ framework that explicitly includes a nonzero $sq$ interaction. 
Interpreted as a hadronic EOS, it effectively represents a maximally stiff limit because additional attractive components (cf. $\sigma^\ast$) are omitted. 
Nevertheless, absent repulsion from the $K^\ast$ channels, we will see that the resulting EOS fails to satisfy observational and experimental constraints.
This setup is not necessarily unrealistic. As an illustrative case, the $\Sigma^+ p$ interaction has been reported to show only limited attraction and a repulsive component in combined experimental and theoretical analyses~\cite{Nanamura2022}.

This paper is organized as follows. 
In Sec.~II, we outline our framework for CSMD. 
Sec.~III presents the numerical results, focusing on the ratio $g_{qK^*}/g_{q\omega}$. Sec.~IV is devoted to the discussion and conclusions.

\section{\label{sec:level2}Color-Spin Molecular Dynamics}
Our basic formulation follows our previous studies \cite{MaruyamaHatsuda2000, Akimura2005, Yasutake2024}. The main advances in the present work are as follows: (i) we include strangeness, (ii) we solve the evolution equations for the internal spin degrees of freedom, and (iii) we consistently treat the associated spin-dependent interactions. The developments corresponding to (ii) and (iii) have already been presented in Refs.~\cite{YasutakeMaruyama2024, MaruyamaYasutake2024}.

We model the $N$-quark system with a (time-dependent) variational ansatz in which each quark is represented by a Gaussian wave packet; the total many-body wave function is the direct product,
\begin{equation}
\begin{split}
\Psi &( {\bf r}_1, {\bf r}_2, ...{\bf r}_N )=\\
&\prod_{i=1}^{N} \frac{1}{(\pi {L_q}^2)^{3/4}}
\exp \left[ -\frac{({\bf r}_i-{\bf R}_i)^2}{2{L_q}^2}
+\frac{i}{\hbar}{\bf P}_i{\bf r}_i
\right]\chi_i,
\end{split}
\label{eq:Gaussian}
\end{equation}
where ${\bf R}_i$ and ${\bf P}_i$ are the position and momentum of the center of each wave packet, respectively, $L_q$ denotes the fixed width of the wave packets, and $\chi_i$ is the internal degree of freedom given by the direct product of flavor ${\chi_i}^f$, color ${\chi_i}^c$, and spin ${\chi_i}^s$. The explicit forms of the time-dependent internal color and spin states are
\begin{eqnarray}
{\chi_i}^c  = \left(
\begin{array}{c}
\cos\alpha_i \;e^{-i\beta_i}\;\cos\theta_i  \\
\sin\alpha_i \;e^{+i\beta_i}\;\cos\theta_i  \\
\sin\theta_i\;e^{i\varphi_i} \\
\end{array}
\right),~~
\chi^S_i = \begin{pmatrix} e^{-i \zeta_i} \cos{\xi_i} \\ e^{i \zeta_i} \sin{\xi_i}  \end{pmatrix},\nonumber\\
\label{eq:colspn}
\end{eqnarray}
where $\alpha_i$, $\beta_i$, $\theta_i$, $\varphi_i$ specify the color state, and $\xi_i$, $\zeta_i$ specify the spin state of each particle.
The flavor composition is determined by minimizing the energy with respect to the flavor fractions. In practice, we solve self-consistently for flavor conversion under charge neutrality and $\beta$ equilibrium, $u+e^- \leftrightarrow d \leftrightarrow s$. For simplicity, we do not include muons in this paper. The present work differs from previous studies in that strangeness is now determined self-consistently together with the color and spin variables.

For energy minimization, we adopt the following Hamiltonian:
\begin{equation}
H = H_0+ V_{\rm Pauli},
\label{eq: H}
\end{equation}
where $V_{\rm Pauli}$ is the phenomenological Pauli potential introduced to reproduce Pauli blocking effects according to our previous studies, and $H_0$ is the conventional Hamiltonian. In this work, we neglect the spurious zero-point energy associated with the cluster center-of-mass motion.

The conventional part of the Hamiltonian takes the standard form
\begin{equation}
H_0=\left<\Psi\left|
\sum_{i=1}^{N}\sqrt{m_i^2+\hat{\textbf{p}}^2_i}+ \hat  V_{\rm C}  + \hat  V_{\rm CS} +  \hat  V_{\rm M}
\right| \Psi \right>. 
\end{equation}
From left to right, the terms are the relativistic kinetic energy, the color-dependent potential, the color- and spin-dependent potential, and the color-independent interaction potential. The constituent quark mass $m_i$ is set as $m_{u,d}= 361.8$ MeV for light quarks, and $m_s= 538.8$ MeV for the strange quark in this study.

As in previous studies, we take the second term to be
\begin{eqnarray}
\hat V_{\rm C}&=&-\frac{1}{2}\sum_{i.j\ne i}^{N}\sum_{a=1}^8\frac{\lambda_i^a\lambda_j^a}{4}\left( \kappa \hat{r}_{ij}-  \frac{\alpha_s}{\hat{r}_{ij}} \right),
\label{eq: Vcol}
\end{eqnarray}
where $\hat{r}_{ij}\equiv |\hat{\bf r}_i-\hat{\bf r}_j|$ is the distance between $i$-th and $j$-th quarks, and $\lambda^a_i$ is the Gell-Mann matrix. The first term is the confining potential, and the second is the one-gluon-exchange (OGE) term. The confinement tension $\kappa$ and the QCD fine structure constant $\alpha_s$ are set as $\kappa=0.75$ GeV~fm$^{-1}$, and $\alpha_s=1.25$~\citep{MaruyamaHatsuda2000}.

The interaction involving the color and spin degrees of freedom is referred to as the color-magnetic interaction:
\begin{eqnarray}
\hat V_{\rm CS} =\frac{1}{2}\sum_{i.j\ne i}^{N}\sum_{a=1}^8 \sum_{b=1}^3 \frac{g_{\rm CS} }{m_i m_j r_{0ij}^2}  \frac{e^{-(\hat r_{ij}/r_{0ij})^2 } }{\hat r_{ij}} \frac{\lambda_i^a \lambda_j^a}{4} \frac{\sigma_i^b \sigma_j^b}{4}, \nonumber \\
 \label{eq:cm}
\end{eqnarray}
where  $\sigma^b_i$ is the Pauli matrix, and \( r_{0ij} \) represents the effective range of the interaction, defined as \( r_{0ij} = 1/(\alpha + \beta \mu_{ij}) \), where \( \mu_{ij} = m_i m_j / (m_i + m_j) \) is the reduced mass. The parameters \( \alpha \) and \( \beta \), which determine the effective range dependence on the reduced mass, are chosen as \( \alpha = 2.1 \, \rm{fm^{-1}} \) and \( \beta = 0.552 \), as provided in Ref.~\cite{Aaron2020}. 
In this method, we do not antisymmetrize the total wave function. As a result, the expectation value of $\lambda_i^a \lambda_j^a$ is underestimated by a factor of four. To compensate for this in the color interaction, we introduce an effective coupling $\kappa_{\mathrm{eff}}$ defined by $\kappa_{\mathrm{eff}} = 4\kappa$~\cite{MaruyamaHatsuda2000}.
A similar issue arises in the quark-spin (color-magnetic) interaction. However, the corresponding coupling constant is calibrated to reproduce the experimentally observed baryon masses (see below). We therefore regard the fitted value of $g_{\rm CS}$ as effectively incorporating the missing antisymmetrization, and we set $g_{\rm CS}=0.34$.

Following previous studies, we introduce a color-independent quark-meson-coupling interaction $V_M$. This implementation requires mapping realistic baryon-baryon forces onto effective quark-quark interactions. The $\sigma$-$\omega$-$\rho$-$\phi$ model is widely used in studies of hadronic matter with strangeness, and in our earlier work we introduced the corresponding $qq$ and $ss$ channels \cite{Akimura2005}. 
In this paper, we extend the setup to a $\sigma$-$\omega$-$\rho$-$\phi$-$K^\ast$ framework that explicitly includes a nonzero $sq$ interaction. The mapping, however, is nontrivial. Derivations starting from realistic baryonic interactions in few-body systems with hyperons indicate that an $sq$ (strange-light) channel---absent in our previous implementations---naturally arises \cite{Yamamoto2022}. The $sq$ channel is the main difference from our previous studies, and we focus on its impact in this work. The interaction takes the following form:

\begin{widetext}
\begin{eqnarray}
\hat V_{\rm M}= \frac{1}{2}\sum_{i}^{N} \left[
\frac{g_{q\omega}^2C_\omega}{4 \pi}  \left(  \sum^N_{j \ne i} \frac{e^ {- \mu _\omega \hat{r}_{ij}}}{\hat{r}_{ij}} \right)^{1+\epsilon_\omega} \right. 
\hspace{-7mm}
&-&  \frac{g_{q\sigma}^2C_\sigma}{4 \pi}  \left(  \sum^N_{j \ne i} \frac{e^ {- \mu _\sigma \hat{r}_{ij}}}{\hat{r}_{ij}} \right)^{1+\epsilon_\sigma}
\hspace{-7mm}
+  \frac{g_{q\rho}^2 C_\rho}{4 \pi}  \left( \sum^N_{j \ne i} \frac{e^ {- \mu _{ \rho } \hat{r}_{ij}}}{\hat{r}_{ij}} \frac{\tau_i^3 \tau_j^3}{4} \right)^{1+\epsilon_\rho}  
\nonumber \\
&& + \left.  \frac{g_{q\phi}^2C_\phi}{4 \pi}  \left(  \sum^N_{j \ne i} \frac{e^ {- \mu _\phi \hat{r}_{ij}}}{\hat{r}_{ij}} \right)^{1+\epsilon_\phi} \hspace{-7mm}
+ \frac{g_{qK^*}^2C_{K^*}}{4 \pi}  \left(  \sum^N_{j \ne i} \frac{e^ {- \mu _{K^*} \hat{r}_{ij}}}{\hat{r}_{ij}} \right)^{1+\epsilon_{K^*}}
\right]. \nonumber\\
~~~
\label{eq: Vmeson}
\end{eqnarray}
\end{widetext}
\red{The first three terms correspond, respectively, to the $\omega$, $\sigma$, and $\rho$ meson exchanges mapped onto effective interactions in the light-quark sector ($u,d$), where $\tau_i^3$ denotes the isospin operator for quark $i$. In the present implementation, these terms are not assigned to $s$ quarks~\cite{Stone2007}. The fourth term accounts for $\phi$ meson exchange mapped onto $ss$-quark interactions, assumed to be repulsive~\cite{Akimura2005,Wu2025}.}
\red{The final term represents the quark-$K^*$ coupling term, i.e., the light-strange ($qs$) interaction, which was absent from our earlier work. Thus the flavor assignments are $(qq)$ for $\sigma,\omega,\rho$, $(ss)$ for $\phi$, and $(qs)$ for $K^*$.}
For all interaction terms, we introduce a nonlinear effect represented by a multiplicative factor $1+\epsilon_i$ 
(with $i \in \{\sigma, \omega, \rho, \phi, K^*\}$). When $\epsilon_i = 0$, the model reduces to purely two-body interactions. By introducing $\epsilon_i$, many-body effects are effectively incorporated into the interaction terms. As will be shown below, this effective many-body contribution is required, within the present framework, to support $\sim 2 M_\odot$ neutron stars. \red{To avoid an excessive number of free parameters, we fix these nonlinearities and keep them unchanged throughout the $g_{qK^*}$ scan.} We set $\epsilon_{\omega}=\epsilon_{\phi}=\epsilon_{K^*}=0.25$ and $\epsilon_{\rho}=-\epsilon_{\sigma}=0.20$. These parameter choices imply that the deviation from a purely two-body interaction remains modest.

For the coupling constants, we use $g_{q\omega}=5.07$, $g_{q\sigma}=3.68$, and $g_{q\rho}=5.15$. These values are based on the simple assumption $g_{q i}=g_{N i}/3$. Following previous studies, we set $g_{s\phi}=\sqrt{2} g_{q\omega}$~\cite{Akimura2005}. 

As discussed, $g_{qK^*}$ is newly introduced here; its coupling strength is not fixed, and mapping realistic baryonic interactions onto a $qs$ channel is inherently nontrivial. Accordingly, we sample five choices, $g_{qK^*} \in\{0, 0.1, 0.3, 0.4, 0.5\}\times g_{q\omega}$, to assess the impact of the strange-light quark coupling. As in our previous papers, we introduce $C_i$ to render the coupling constant $g_{qi}$ dimensionless, setting $C_i=1/(1+\epsilon_i)$.
For the parameters $\mu_i$ appearing in the numerators of the interaction terms, we use the masses of the mediator mesons:
$\mu_\omega=782.6~\mathrm{MeV}$, $\mu_\sigma=400~\mathrm{MeV}$, $\mu_\rho=770~\mathrm{MeV}$, and $\mu_{K^*}=892~\mathrm{MeV} $.
These interaction choices may appear ad hoc; indeed, to some extent they are, for several reasons. First, in three-body systems---i.e., when obtaining baryon masses---the dominant contributions arise from color- and spin-dependent terms, which makes the optimization of these color-independent potentials difficult in few-body settings. Second, in the high-density regime we constrain the EOS primarily through the mass-radius ($MR$) relation from astronomical observations; however, that constraint is not yet decisive enough to determine many channels (e.g., $\sigma^*$, $\delta$, $\kappa$). We therefore focus on the impact of the repulsive vector $K^*$ channel in this work.

As discussed previously, the interaction kernels in Eqs. (1) and (4) are computed through a double convolution with the quark Gaussian wave packets. In line with our earlier studies, the effective range of the color-independent potentials is controlled by the packet-width parameter $L$ in Eq. (1) for each channel.
We set the effective widths as $L_\omega=L_\phi=L_{K^*}=0.92$ fm, $L_\sigma=1.20$ fm, and $L_\rho=1.35$ fm for the respective meson-exchange terms. \red{The somewhat larger value adopted for the $\rho$ channel is a phenomenological choice used to regulate the isovector sector and obtain reasonable symmetry-energy properties in the light-quark EOS; it is then kept fixed in the subsequent $g_{qK^*}$ scan.}

Rather than performing explicit antisymmetrization, we incorporate fermionic exclusion via an effective Pauli potential $V_{\mathrm{Pauli}}$, which provides a repulsion between quarks with identical internal quantum numbers (flavor, color, spin), denoted $\chi_i$:
\begin{eqnarray}
V_{\rm Pauli}=   \frac{1}{2}\sum_{i,j \ne i}^{N} &&
\frac{C_p}{(q_0 p_0)^3}\exp{\left[ -\frac{({\bf R}_i-{\bf R}_j)^2}{2q_0^2}\right]} \nonumber\\
&&\times \exp{\left[ -\frac{({\bf P}_i-{\bf P}_j)^2}{2p_0^2}\right]} \delta_{\chi_i,\chi_j} 
\label{eq: Pauli}
\end{eqnarray}
We adopt $q_0=3~\mathrm{fm}, p_0=150~\mathrm{MeV}$, and $C_p=190~\mathrm{MeV}$, which reproduce the zero-temperature relativistic kinetic energy of free fermions. 
The system evolves according to the Euler--Lagrange equations for the generalized coordinates 
\{${\bf R}_i$, ${\bf P}_i$, $\alpha_i$, $\beta_i$, $\theta_i$, $\varphi_i$, $\xi_i$, $\zeta_i$ \}.
The resulting equations of motion are
\begin{eqnarray}
\dot {\bf R}_i
&=&\frac{\partial H}{\partial {\bf P}_i}, \ \ \ \ 
\dot {\bf P}_i
=-\frac{\partial H}{\partial {\bf R}_i}, \label{eq: rp_evo}\\
\dot\alpha_i
&=&\frac{1}{2\hbar\sin2\alpha_i\;\cos^2\theta_i}\frac{\partial H}{\partial\beta_i}
   -\frac{\cos2\alpha_i}{2\hbar\sin2\alpha_i\;\cos^2\theta_i}\frac{\partial H}{\partial\varphi_i}, \nonumber \\
~~~\\
\dot\beta_i
&=&-\frac{1}{2\hbar\sin2\alpha_i\;\cos^2\theta_i}\frac{\partial H}{\partial\alpha_i},\\
\dot\theta_i
&=&\frac{1}{ \hbar\sin{2\theta_i}}\frac{\partial H}{\partial\varphi_i},\\
\dot\varphi_i
&=&-\frac{1}{ \hbar\sin{2\theta_i} }\frac{\partial H}{\partial\theta_i}
  +\frac{\cos2\alpha_i}{2\hbar\sin2\alpha_i\;\cos^2\theta_i}\frac{\partial H}{\partial\alpha_i}, \\
\dot\xi_i
&=&\frac{1}{2\hbar\sin2\xi_i}\frac{\partial H}{\partial\zeta_i},  \\
 \dot\zeta_i
&=&-\frac{1}{2\hbar\sin2\xi_i}\frac{\partial H}{\partial\xi_i}.
\label{eq: betaS}
\end{eqnarray}

In the calculations, all quarks are initially distributed randomly, with zero momenta, in a box with periodic boundary conditions.
The ground state (matter at zero temperature) is obtained by the frictional cooling~\cite{MaruyamaHatsuda2000, Akimura2005}.
For this purpose, we solve damped equations of motion instead of Eqs.~(\ref{eq: rp_evo})--(\ref{eq: betaS}),
\begin{eqnarray}
& & \tilde{\dot{v}}_i=\dot {v}_i  +\mu_v \frac{\partial H}{\partial {v}_{i}} \label{eq: rfric}, 
\end{eqnarray}
with $v_i\in\{{\bf R}_i,{\bf P}_i,\alpha_i,\beta_i,\theta_i,\varphi_i,\xi_i,\zeta_i\}$; $\mu_v$ is the corresponding damping coefficient set as $\mu_R=-0.00002$, $\mu_P=-0.02$, and $\mu_{\xi}=\mu_{\zeta}=-0.1$.
If any single damping parameter is activated, the total energy for the system decreases, leading to optimization of all variables. Therefore, it is not necessary to introduce damping for every variable; instead, one should choose the most efficient parameters while avoiding trapping in local energy minima. This is why we do not apply damping parameters to the color degrees of freedom.

Based on the above framework, we calibrate the baseline model by (i) computing finite three-quark systems (baryons) and confronting the results with experimental masses, and (ii) computing infinite NS matter with periodic boundary conditions and comparing with observational constraints. We then explore several values of $g_{qK^*}$ around that baseline. However, one caveat is in order. Our calculations are restricted to two-body interactions; genuine three-body forces are not included explicitly, and many-body effects enter only through the effective nonlinear factors $\epsilon_i$. Beyond this nonlinearity in spatial (distance) correlations encoded by $\epsilon_i$, a proper treatment of many-body effects also requires an appropriate handling of internal degrees of freedom such as color and spin. In this context, traditional few-body constituent-quark models have successfully derived baryon masses and multiquark resonances by explicitly computing many-body correlations in color and spin. However, applying the same machinery directly to MD simulations is not practical from a computational-cost standpoint. Accordingly, we incorporate effective spin-color correlations, by which the many-body correlations are mapped onto averaged two-body ones. For baryon masses, we employ effective spin-color correlations in the $N=3$ case, whereas for EOS we adopt the large-$N$ limit. Further details are given in Appendixes A and B. We show the baryon-mass results including strangeness in Table~\ref{tab:mass}, compared with experiment.
From these masses, the primary constraints are on the coupling constant of the color-magnetic interaction $g_{\rm CS}$, the strange-quark constituent mass $m_s$, and the effective ranges $L_i$.

\begin{center}
\begin{table}[t]
   \begin{tabular}{|c|c|c|c|c|c|c|c|}
  \hline
   $B$        & N, P     & $\Delta$    & $\Lambda$   & $\Sigma$     & $\Sigma^*$   & $\Xi$    &  $\Xi^*$     \\
  \hline \hline
  $M_B$ [MeV]  & 938 & 1233 & (1115) & 1177 & 1379 & 1328 & 1531 \\
  \hline
Expt. [MeV] & 938   & 1232  & 1115  & 1189 & 1382 & 1315 & 1532  \\
  \hline
  \end{tabular}
  \caption{
Baryon masses from our CSMD ($M_B$) compared with experimental values (Expt.). 
As detailed in Appendixes A and B, we do not compute the $\Lambda$ mass explicitly within CSMD. Rather, we infer it using the correspondence $M_\Lambda - M_N \approx m_s - m$ based on a simple constituent-quark model, which we adopt to fix the strange-light mass splitting. For this reason, we enclose $M_\Lambda$ in parentheses.}
\label{tab:mass}
\end{table}
\end{center}

\section{RESULTS}

\subsection{EQUATION OF STATE }
\begin{figure}[tb]
\includegraphics[width=0.47\textwidth]{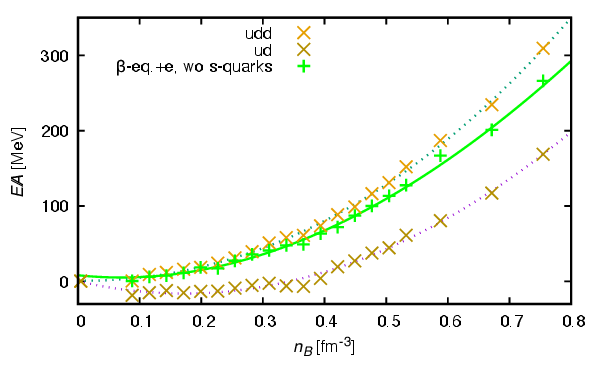}
\caption{\label{fig: eos} 
Density dependence of the energy per baryon $E/A$ for $ud$-matter and $udd$-matter.
Each dot denotes a CSMD data point; thin dotted lines indicate the corresponding regression fits.
The point labeled ``$\beta$-eq.+e" gives $E/A$ under charge neutrality and $\beta$-equilibrium, including the electron contribution.}
\end{figure}

First, excluding strangeness and taking the large-$N$ limit, we obtain the energy per nucleon as shown in Fig.~\ref{fig: eos}. Here the flavor degrees of freedom are fixed. \red{In this limit, $udd$ matter serves as a proxy for neutron-like matter, whereas $ud$ matter serves as a proxy for purely symmetric matter.} The figure additionally shows $\beta$-equilibrated matter without strangeness, with flavor conversion taken into account. The electron contribution is included in that curve. We exclude the low-density regime ($< 0.05~\mathrm{fm}^{-3}$), where finite-size (box) effects become significant, from the fit. In addition to our CSMD results, the figure includes a fit based on the conventional Taylor expansion given below. Although current nuclear-physics experiments do not tightly constrain terms up to third order, we nevertheless perform a fit including terms up to third order in the normalized density $x$. 
\begin{eqnarray}
(E/A)_{ud}(n_B) &=& S_0 + \frac 1 2 K_0 x^2 + \frac 1 6 Q_0 x^3, \\
\label{eq: ud}
(E/A)_{udd}(n_B)&=&  S + Lx + \frac 1 2 K x^2 + \frac 1 6 Q x^3, 
\label{eq: udd}
\end{eqnarray}
where $x \equiv \frac{n_B-n_0}{3n_0}$ with the nuclear saturation density $n_0$.
In general, these curves do not pass through the origin; however, the behavior near the origin is improved by calculations for a single baryon ($N=3$ case) in a box.
Table~\ref{tab:satulation} lists these parameters. As noted above, the third-order terms are provided for reference and are shown in parentheses. The quantity $J$ in the table denotes  the zeroth-order offset $J \equiv S-S_0$, i.e., the symmetry energy. 
The value of $L$ may appear large compared with typical fiducial values~\cite{Vinas2014,Li2019}. However, the possibility of a large $L$ has also been discussed~\cite{Danielewicz2017,Estee2021}. We therefore confine the present analysis to interactions consistent with the parameter set summarized in this table and focus on the impact of the $qK^\ast$ interaction, treating it as the only free parameter.
\red{We do not attempt a quantitative comparison with the chiral effective field theory (chiral EFT) band for neutron matter in the present paper; nevertheless, such a comparison could provide additional guidance in choosing the $\epsilon_i$ parameters associated with many-body effects.}

\begin{center}
\begin{table}[tb]
  \begin{tabular}{|cccccccc|}
     \hline
     $n_0$ & $J$ & $L$ & $K$ & $Q$ & $S_0$ & $K_0$ & $Q_0$  \\
    $[\rm fm^{-3}]$ & [MeV] & [MeV]& [MeV] & [MeV]  & [MeV] & [MeV] & [MeV]  \\
    \hline \hline 
    0.172 & 30.3  & 79.1 & (242) & (67.3) &-16.1 & 256 & (-11.4) \\
    \hline
  \end{tabular}
   \caption{The saturation properties obtained by CSMD calculations. }
  \label{tab:satulation}
\end{table}
\end{center}

\begin{figure}[tb]
\includegraphics[width=0.47\textwidth]{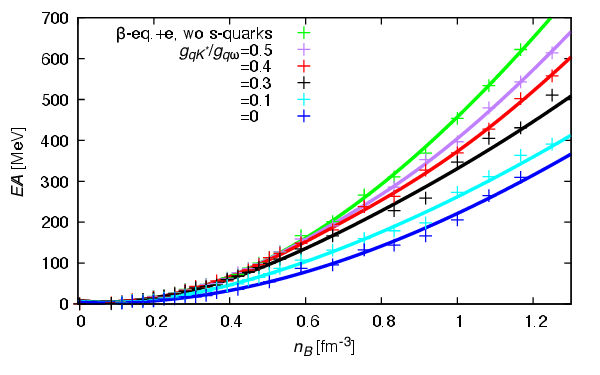}
\caption{\label{fig: eosS} Same as Fig.~\ref{fig: eos}, but with strangeness included for different values of $g_{qK^*}$ scaled relative to $g_{q\omega}$. All data points include the electron contribution under charge-neutral, $\beta$-equilibrated conditions. For reference, the figure also displays 
$E/A$ without $s$ quarks under charge neutrality and $\beta$-equilibrium, including the electron contribution (denoted ``$\beta$-eq.+e, without $s$ quarks").}
\end{figure}

Next, in Fig.\ref{fig: eosS} we consider matter that includes strangeness and impose charge neutrality and $\beta$-equilibrium conditions, allowing the flavor conversion, $u + e^- \leftrightarrow d \leftrightarrow s$. In this figure, we include the electron contribution and show five choices for the $qK^*$-interaction strength relative to the $q\omega$ interaction: $g_{qK^*}/g_{q\omega}=0$, 0.1, 0.3, 0.4, and 0.5. Accompanied by the onset of strangeness, the data points indicate nontrivial behavior that deviates from a simple third-order Taylor expansion.
Hence, we adopt the following fitting function:
\begin{eqnarray}
(E/A)_{\beta}(n_B)&=&  S_{\beta} + L_{\beta}x + \frac 1 2 K_{\beta} x^2 + \frac 1 6 Q_{\beta} x^3, 
\label{eq: beta}\\
(E/A)_{uds}(n_B)&=& (E/A)_{\beta}(n_B) \cdot \Big\{ 1+ a_1 \sigma(a_2 y-a_3)  \Big\} , \nonumber \\
\label{eq: uds}
\end{eqnarray}
where $\sigma(x)$ is a sigmoid function, and $y \equiv n_B/\bar n$ with $\bar n = 0.16~\mathrm{fm}^{-3}$.
We emphasize that $\bar n$ is not the saturation density; it is introduced solely as a typical density for normalization.
Dividing out the overall factor of 3 yields the energy per quark $E/Q$.
Eqs.~(\ref{eq: beta}) and (\ref{eq: uds}) are adopted as an interpolation formula for constructing a continuous EOS for the TOV calculation. Equation~(\ref{eq: beta}) provides a smooth baseline for the $\beta$-equilibrated matter without strangeness, while Eq.~(\ref{eq: uds}) augments it with a sigmoid factor to capture the onset of strangeness.
The fitted parameters for Eq.~(\ref{eq: beta}) are listed in Table~\ref{tab:fitting-beta}, while those for Eq.~(\ref{eq: uds}), together with the onset density $n_c$ at which $s$ quarks appear in the matter, are listed in Table~\ref{tab:fitting-uds}.
In many existing EOS that include strangeness---whether through hyperons or kaons---the strangeness onset occurs at densities $\gtrsim 2n_0$. 
\red{The $g_{qK^*}=0$ case corresponds to the conventional no-$K^*$ choice, in which the $qK^*$ channel is omitted. As $g_{qK^*}$ increases across the sampled set, the onset density shifts upward. Relative to the commonly quoted scale, the $0.4$ and $0.5$ cases come closer to that range, whereas the smaller-coupling cases are retained mainly as reference points for illustrating the sensitivity to the strange-light interaction.} While the reliability of infinite-matter predictions calibrated solely by finite-system constraints (e.g. experimental data of heavy-ion collisions) is nontrivial, we use
\begin{eqnarray}
g_{qK^*} /g_{q\omega} \gtrsim 0.4,
\label{eq: g_qs_cnst1}
\end{eqnarray}
as a practical benchmark below.

\begin{table}[tb]
\centering
  \begin{tabular}{|cccc|}
    \hline
    $S_{\beta}$ & $L_{\beta}$ & $K_{\beta}$ & $Q_{\beta}$ \\
    \multicolumn{1}{|c}{[MeV]} &
    \multicolumn{1}{c}{[MeV]} &
    \multicolumn{1}{c}{[MeV]} &
    \multicolumn{1}{c|}{[MeV]} \\
    \hline\hline
    11.3143  &  57.4012  & 265.169 & -55.8235   \\
    \hline
  \end{tabular}
  \caption{Optimized parameters in Eq.~(\ref{eq: beta}) for $\beta$-equilibrated matter without strangeness.}
  \label{tab:fitting-beta}
\end{table}

\begin{table}[tb]
\centering
\begin{tabular}{|c|c|ccc|}
\hline
 $g_{qK^*}/g_{q\omega}$ & $n_c$ & $a_1$ & $a_2$ & $a_3$   \\
 & [fm$^{-3}$] & &  &  \\
\hline \hline
 0    & $< $0.05 &  -0.516137 & 1.19025 & -1.80282  \\
0.1  & 0.15 & -0.487487 & 0.372239  & 0.373212    \\
0.3  & 0.20 & -0.367451 & 0.552446 & 2.34678  \\
0.4  & 0.31 &  -0.205484  & 1.25214 & 5.66601   \\
0.5  & 0.39 &  -0.121334  & 1.72117 & 7.7807   \\
\hline
\end{tabular}
\caption{Optimized parameters in Eq.~(\ref{eq: uds}) and the strangeness onset density $n_c$ for different values of $g_{qK^*}/g_{q\omega}$.}
\label{tab:fitting-uds}
\end{table}

\begin{figure}[tb]
\includegraphics[width=0.49\textwidth]{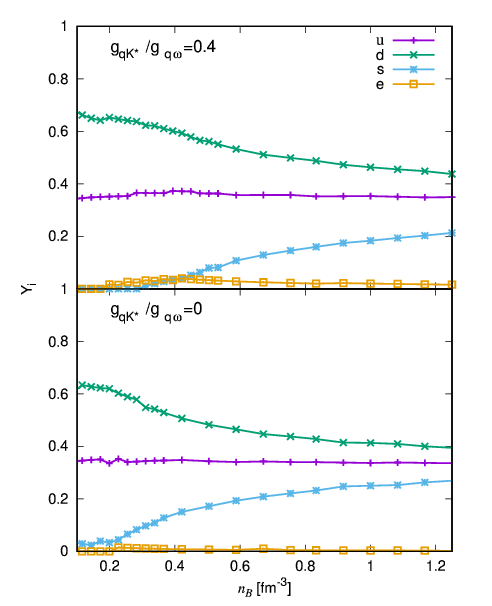} 
\caption{\label{fig:Yi}\textcolor{red}{Quark particle fractions as functions of density for the best-performing sampled case, $g_{qK^*}/g_{q\omega}=0.40$, and for the no-$K^*$ reference case, $g_{qK^*}/g_{q\omega}=0$. Under the adopted clustering criterion, the increase of the $s$-quark fraction should not be interpreted as the emergence of isolated deconfined quarks; as discussed in Sec.~III.C, these quarks are realized as constituents of baryons and larger multiquark configurations.}}
\end{figure}

Quark particle fractions are presented in Fig.~\ref{fig:Yi} for the best-performing sampled case, $g_{qK^*}/g_{q\omega}=0.4$, and for $g_{qK^*}/g_{q\omega}=0$. The remaining cases behave similarly and are therefore not shown to avoid redundancy. With increasing density, the $s$-quark fraction rises; however, larger $g_{qK^*}$ suppresses this fraction. It should be emphasized that these do not appear as isolated quarklike configurations. As we demonstrate below, under the adopted clustering criterion they are realized as constituents of baryons and larger multiquark configurations.

\begin{figure}[tb]
\includegraphics[width=0.49\textwidth]{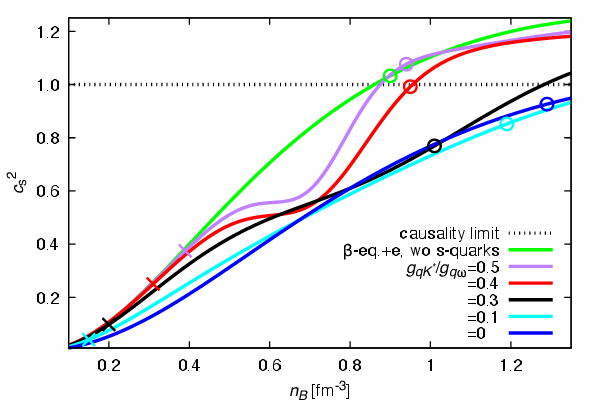}
\caption{\label{fig: cs} 
Squared sound speeds normalized by the speed of light.
The circles indicate the maximum central density corresponding to the maximum NS mass.}
\end{figure}

The corresponding squared sound speeds, normalized by the speed of light, are shown in Fig.~\ref{fig: cs}. 
The figure also marks the onset densities for strangeness by crosses. In the vicinity of these thresholds, some models display a distinctive non-monotonic trend in the sound speed, oscillating up and down as the density increases. Such behavior can resemble a crossover-type hadron-quark conversion (hadron-quark continuity) or the onset of kaons in hadronic matter. Nevertheless, our clustering analysis below suggests that the stable-branch configurations are better interpreted as clustered hadronic matter than as a deconfined quark phase.

Circles indicate the maximum central density reached in each sequence, corresponding to the maximum mass.
The model with $g_{qK^*}/g_{q\omega}=0.5$ and the model without $s$ quarks encounter causality violation before reaching the maximum mass. Consequently, within the sampled set, the sound-speed analysis favors the remaining models:
\begin{eqnarray}
0 \leq g_{qK^*}/g_{q\omega} \lesssim 0.4.
\label{eq: g_qs_cnst2}
\end{eqnarray}
While the present condition does not violate causality within neutron-star interiors, some of our models still violate causality at higher densities. This reflects the fact that our framework includes relativistic effects only partially. 
A fully relativistic formulation in which both the interaction terms and the kinetic energy are treated on the same relativistic footing remains a task for future work.

\subsection{MASS-RADIUS RELATION}

\begin{figure}[tb]
\includegraphics[width=0.49\textwidth]{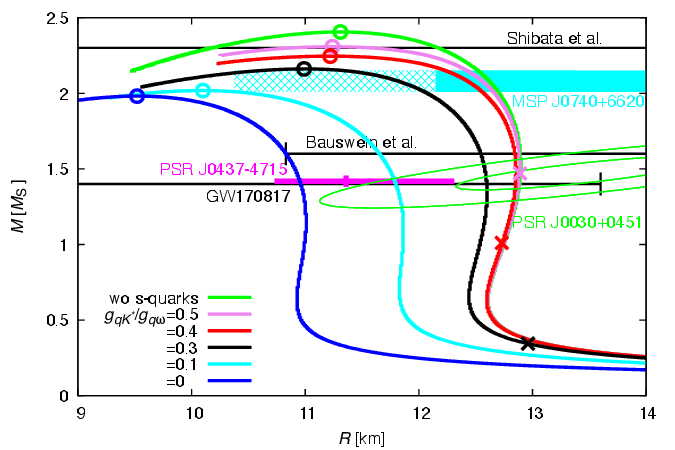}
\caption{\label{fig: MR} Mass-radius (MR) relations obtained from the CSMD results. For each sequence, the maximum mass is marked by a circle and the strangeness onset density by a cross. We also overlay constraints from PSR J0030+0451~\cite{Riley2019,Miller2019}, PSR J0740+6620~\cite{Riley2021,Miller2021}, 
PSR J0437+4715~\cite{Choudhury2024}, GW170817 together with its EM counterpart~\cite{Bauswein2017}, and the numerical-relativity upper mass limit~\cite{Shibata2019}. As for PSR J0030+0451, revised MR inferences have been published by Vinciguerra et al., who report two model-dependent solutions~\cite{Vinciguerra2024}. Taken together, these two solutions span a range that is broadly consistent with the previously quoted $2\sigma$ interval; effectively, they correspond to the two extremes of the earlier inference. See the main text and references for further details.}
\end{figure}

We next examine the neutron-star MR relation computed with CSMD (Fig.~\ref{fig: MR}). 
We use the Baym et al.~\cite{Baym1971} crust EOS for densities $n_B < \tfrac{1}{3}n_0$ and interpolate between that EOS and our model over $\tfrac{1}{3}n_0 \le n_B \le \tfrac{2}{3}n_0$. In the inner crust at low density, nuclear pasta phases appear on length scales of $\sim 10\text{-}30~\mathrm{fm}$ as a result of the competition between Coulomb repulsion and surface tension. A rigorous treatment therefore necessitates including Coulomb interactions and either varying the periodic-box size to identify the optimal morphology or employing a sufficiently large simulation domain~\cite{Caplan2018}. Such calculations are beyond the scope of this work.

In this figure, circles denote the maximum mass along each sequence, while crosses indicate the corresponding strangeness onset.
For comparison, we overlay constraints from PSR J0437+4715~\cite{Choudhury2024}, PSR J0030+0451~\cite{Riley2019, Miller2019}, PSR J0740+6620~\cite{Riley2021,Miller2021}, GW170817~\cite{Abbott2017} and its electromagnetic (EM) counterpart~\cite{Bauswein2017}, as well as the numerical-relativity upper bound on the maximum NS mass~\cite{Shibata2019}.
For PSR J0030+0451, revised mass-radius inferences have been published, reporting two model-dependent solutions:
$M=1.40^{+0.13}_{-0.12} M_\odot, R=11.71^{+0.88}_{-0.83} \mathrm{km}$, and
$M=1.70^{+0.18}_{-0.19} M_\odot, R=14.44^{+0.88}_{-1.05} \mathrm{km}$~\cite{Vinciguerra2024}.
As both fall at the two extremes within the 2$\sigma$ bounds of Ref.~\cite{Miller2019}, we do not display them in the figure to avoid clutter.
In addition to these constraints, nucleosynthesis calculations in numerical relativity for black hole-neutron star merger ejecta favor a DD2-like EOS \cite{Wanajo2024}, which yields $R=13.2~\mathrm{km}$ at $M=1.4 M_\odot$ \cite{DD2}. We therefore do not regard the current radius constraints as decisive and instead focus on the impact of the $qs$ interaction---interpreted as arising from $qK^\ast$ coupling---on NS radii.

All sampled models achieve $M_{\max}>2M_\odot$. This indicates that, within our framework, the nonlinear components $\epsilon_i$, representing effective many-body physics, are required to resolve the hyperon puzzle. By contrast, the predicted radii vary appreciably depending on the coupling strength $g_{qs}$. In contrast, the results reported here were obtained while keeping the nucleonic symmetry energy fixed.
Instead, we treat the strength of $g_{qK^*}$ as a free parameter. 
MR relations with $g_{qK^*} /g_{q\omega}  \leq 0.10$ fail to satisfy the PSR J0740+6620 constraint, while those with $g_{qK^*} /g_{q\omega} \gtrsim 0.50$ violate the numerical-relativity upper limit. Consequently, within the sampled set, astrophysical data favor 
\begin{eqnarray}
0.10 <g_{qK^*} /g_{q\omega} <0.50.
\label{eq: g_qs_cnst3}
\end{eqnarray}
Table~\ref{tab: MmaxLam} summarizes these results. The table also reports the tidal deformability at $1.4 M_\odot$, $\Lambda_{1.4M_\odot}$; none of the sampled models are ruled out by this observable.

In summary, among the sampled values explored, one sampled case provides the best overall consistency with all three criteria---(i) strangeness onset at $\sim 2n_0$, (ii) the causality condition, and (iii) the mass-radius (MR) relation, i.e. Eqs.~(\ref{eq: g_qs_cnst1})-(\ref{eq: g_qs_cnst3})---namely the best-performing sampled point,
\begin{eqnarray}
g_{qK^*}/g_{q\omega}=0.40.
\label{eq:g_qs_cnst}
\end{eqnarray}
\red{We emphasize that this value is the best-performing one within the discrete sampled set considered in this work, not a unique best-fit parameter.}
\red{It should be regarded as provisional, since the continuously allowed interval around this value has not yet been mapped out and could be modified once additional physics is incorporated, for example an attractive $qK$ channel or a fully relativistic extension in which both the interaction terms and the kinetic energy are treated on the same relativistic footing.}
\red{Future neutron-star observations, especially if radius determinations converge more decisively toward either smaller or larger values, should provide a stronger empirical restriction on the allowed range.}

\begin{center}
\begin{table}[bt]
\begin{tabular}{|c|ccc|}
\hline
 $g_{qK^*}/g_{q\omega}$ & $M_{\rm max}$ [$M_\odot$] & $n_{\rm B, max}$ [fm$^{-3}$] &$\Lambda_{1.4M_\odot}$ \\
\hline \hline
 0      & 2.02 & 1.29   & 214 \\
0.10  & 2.02 & 1.19   & 333 \\
0.30  & 2.16 & 1.01  & 498 \\
0.40  & 2.25 & 0.952   & 588 \\
0.50  & 2.31 & 0.935 & 596  \\
without $s$ quarks            & 2.39 & 0.905 & 603  \\
\hline
\end{tabular}
\caption{Maximum mass $M_{\rm max}$, the corresponding central density $n_{\rm B, max}$, and  
the tidal deformability at $1.4 M_\odot$, $\Lambda_{1.4M_{\odot}}$
for each model.}
\label{tab: MmaxLam}
\end{table}
\end{center}

\newpage
\subsection{CLUSTERING}

\begin{figure*}[bth]
\begin{flushleft}
(a) \vspace{-1.3cm}
\end{flushleft}
\includegraphics[width=0.97\textwidth]{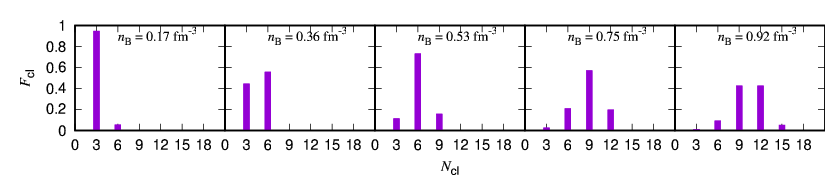} \\
\begin{flushleft}
(b) \vspace{-1.3cm}
\end{flushleft}
\includegraphics[width=0.96\textwidth]{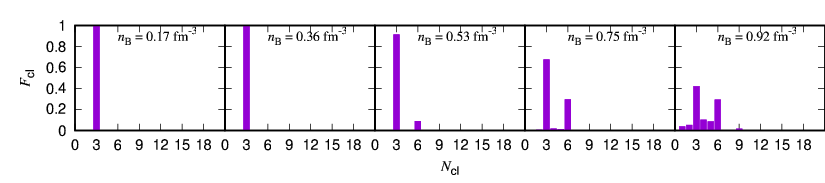} \\
\begin{flushleft}
(c) \vspace{-1.3cm}
\end{flushleft}
\includegraphics[width=0.96\textwidth]{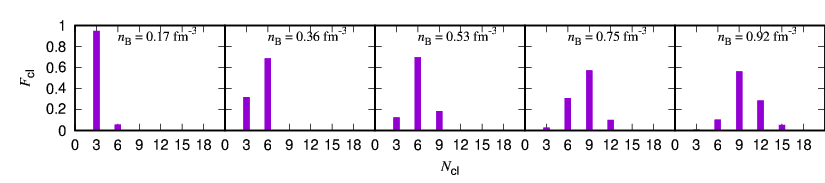} \\
\begin{flushleft}
(d) \vspace{-1.3cm}
\end{flushleft}
\includegraphics[width=0.96\textwidth]{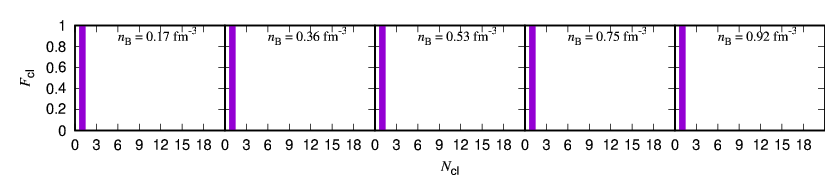}%
\caption{\label{fig:clst_cmag} Distribution of cluster size $N_{\rm cl}$ (number of quarks per cluster) and quark number fraction $F_{\rm cl}$ as functions of density. Panel (a) shows the best-performing sampled case, $g_{qK^*}/g_{q\omega}=0.40$, panel (b) is the same as (a) but without the color-magnetic interaction, panel (c) is without strangeness but includes the color-magnetic interaction, and panel (d) is a random configuration for comparison.
}
\end{figure*}

\begin{figure*}[tb]
(a)~~~~~~~~~~~~~~~~~~~~~~~~~~~~~~~~~~~~~(b)~~~~~~~~~~~~~~~~~~~~~~~~~~~~~~~~~~~~~(c)~~~~~~~~~~~~~~~~~~~~~~~~~~~~~~~~~~~~~(d)\\
\includegraphics[width=0.24\textwidth]{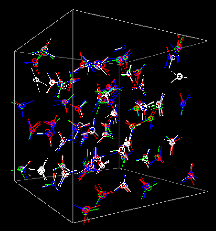}
\hspace{-2mm}
\includegraphics[width=0.2395\textwidth]{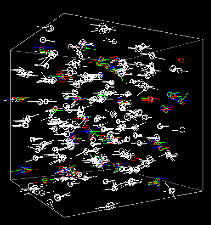}
\hspace{-2mm}
\includegraphics[width=0.2515\textwidth]{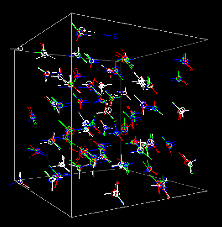}
\hspace{-2mm}
\includegraphics[width=0.237\textwidth]{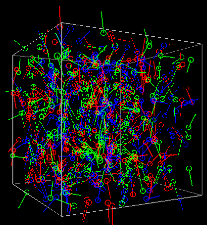}
\caption{\label{fig: disp} Quark configurations at baryon density $n_B=0.92~\mathrm{fm}^{-3}$. For visualization, a short rod affixed to each particle denotes the spin orientation. Panel (a) shows the best-performing sampled case, $g_{qK^*}/g_{q\omega}=0.40$, panel (b) is the same as (a) but without the color-magnetic interaction, panel (c) is the same as (a) but without strangeness, and panel (d) is the corresponding random configuration.} 
\end{figure*}

In this section, we examine how confinement manifests itself in the high-density configurations. As demonstrated below, the color-magnetic interaction is pivotal to this behavior.
The following cluster analysis is based on an operational criterion: if one quark lies within a distance $d_{\rm cluster}$ of another quark, the two are assigned to the same cluster. 
In this work we adopt $d_{\rm cluster}=0.15$ fm as the clustering threshold.
\red{For $N_{\rm cl}=6$, the physically relevant distinction is between a dibaryon-like molecular configuration and a compact sexaquark-like multiquark state, and the corresponding characteristic size scale is approximately $1~\mathrm{fm}$. Since the effective widths of the interaction kernels are also approximately $1~\mathrm{fm}$, our choice of $d_{\rm cluster}$ is intended to tag only configurations in which the quarks nearly share a common spatial center, rather than two more weakly associated baryonic clusters.}
\red{To gauge the sensitivity to this choice, Appendix C compares the cases with and without the color-magnetic interaction for $d_{\rm cluster}=0.10$ and $0.20$ fm.}

Fig.~\ref{fig:clst_cmag} presents the distribution of cluster sizes $N_{\rm cl}$ (the number of quarks per cluster) as a histogram, plotted with quark number fraction $F_{\rm cl}$. Note that this distribution depends quantitatively on the adopted distance threshold, so we use it mainly to identify qualitative trends rather than to assign unique hadron species. The plot simply records, for each quark, which cluster it belongs to; it is not normalized by cluster size. Consequently, a baryon with $N_{\rm cl}=3$ and an exotic hadron with $N_{\rm cl}=15$ contribute differently: the latter yields a fraction $F_{\rm cl}$ five times larger. 
Here, the density dependence is shown up to $0.92~\mathrm{fm}^{-3}$, which is close to the maximum density reached in any of our cases.

Panel (a) corresponds to the best-performing sampled case, $g_{qK^*}/g_{q\omega}=0.40$, and shows that the clusters increase in size with increasing density while continuing to satisfy $N_{\rm cl}=3n$ ($n\in\mathbb{N}$) under the adopted criterion. This result suggests that isolated quarklike configurations, i.e., $N_{\rm cl}=1$, do not appear even at high density under this criterion; instead, multiquark clusters with $N_{\rm cl}$ up to 15 do appear. To investigate the physical origin of this behavior, we analyze panels (b), (c), and (d).

As in panel (a), panel (b) shows the distribution of cluster sizes, but with the color-magnetic interaction switched off. Compared with panel (a), the cluster-size fraction $F_{\rm cl}$ is shifted toward smaller $N_{\rm cl}$ at all densities, and the condition $N_{\rm cl}=3n$ is no longer satisfied at high density. In particular, clusters with $N_{\rm cl}=1$---i.e., isolated quarklike configurations---appear at the highest density shown in this panel, $n_B = 0.92~\mathrm{fm}^{-3}$. This behavior is consistent with our previous results \cite{Yasutake2024}.

Panel (c) shows the case without strangeness but including the color-magnetic interaction. Although the cluster-size fraction $F_{\rm cl}$ is again shifted toward smaller $N_{\rm cl}$ at all densities compared with panel (a), isolated quarklike configurations do not emerge under the adopted criterion.

Panel (d) is included to examine whether increased density by itself leads to clustering. This configuration is obtained by placing the particles at random positions. Without correlated color organization, clustering does not occur because the mean distance clearly exceeds the critical distance.

From these panels in Fig.~\ref{fig:clst_cmag}, we infer that, within the adopted clustering criterion, isolated quarklike configurations do not appear along the stable branch when the color-magnetic interaction is taken into account, regardless of the presence of strangeness. Instead, larger multiquark clusters are favored. However, as discussed below, because of our treatment of spin configurations, the present calculation does not provide a physically exact assignment of specific compositions to individual clusters. For example, we cannot identify whether a given cluster is neutron or proton. These limitations should be regarded as artifacts of the coherent-state-based formulation. Accordingly, we refrain from listing the composition of each cluster. 

To determine whether the resulting clusters are genuinely multiquark states rather than hadronic molecules, it is more informative to inspect their configurations directly. 
The quark configurations for the best-performing sampled case, $g_{qK^*}/g_{q\omega} = 0.40$, are shown in panel (a) of Fig.~\ref{fig: disp}, where the baryon density is $n_B = 0.92~\mathrm{fm}^{-3}$. This density is close to the maximum density inside stable NSs (see Table~\ref{tab: MmaxLam}).
Colors denote the proximity of each quark's color state to R, G, or B; triplets within the same cluster whose total color is singlet (color neutral) are rendered in white. Note that not all quarks are shown in white, even though---as in Fig.~\ref{fig:clst_cmag}---all quarks are arranged into clusters of size $N_{\rm cl}=3n$ ($n\in\mathbb{N}$). This has a physical basis: each cluster is not merely a collection of baryons; rather, colors are mutually correlated within it, and the cluster as a whole is color singlet. For clarity, the spin orientations are represented by short rods attached to each particle in this figure. The orientations form a Y-shape within each cluster. This arises because (i) color and spin are modeled as continuous classical variables in Eq.~(\ref{eq:colspn}), rather than quantized ones, and (ii) we work in the large-$N$ limit of spins, which renders the spin strength independent of direction, as in Eqs.~(\ref{eq: spinup}) and (\ref{eq: spindown}) in Appendix A. Owing to the mutual spin correlations and their coupling to color through color-magnetic interactions, the energetically favored configuration in each cluster is an isotropic, Y-shaped arrangement. 
Due to this arrangement, one cannot simply identify $udd-$matter with neutrons or $ud-$matter with protons. 
Therefore, for exotic hadrons as well, it remains uncertain at present how realistic the internal quark configurations actually are.

For comparison, we present in panel (b) the configurations obtained when color-magnetic interactions are switched off. The spins largely retain their initial horizontal orientations.

Panel (c) corresponds to the case in panel (a) but without strangeness. Again, the Y-shaped spin configuration emerges. Compared with panel (b), color-singlet combinations among neighboring three quarks appear less frequently; hence, we conclude that spin and color degrees of freedom are entangled within each cluster via the color magnetic interaction.

Panel (d) shows the configuration obtained by placing the particles at random positions, corresponding to panel (d) in Fig.\ref{fig:clst_cmag}.
As is evident from the figure, each quark is distributed separately and does not form clusters.

Within the scope of our model, high-density conditions such as those inside NSs favor the emergence of hadronlike multiquark clusters of typical size $\lesssim 1~\mathrm{fm}$. Such scenarios have been contemplated in the literature (e.g. sexaquarks~\cite{Shahrbaf2022} and strangeons~\cite{Xia2025}). The novelty of the present work is that, by implementing strangeness and, critically, color-magnetic interactions in molecular-dynamics simulations, we obtain the self-consistent formation of multiquark clusters. Among these ingredients, the color-magnetic interaction plays the central role. \red{In the present analysis, however, a large cluster size is not by itself used as a definition of deconfinement; rather, it is an operational indicator of spatial connectivity under the adopted criterion and is more naturally interpreted as suggestive of hadronlike exotic multiquark or multibaryon structures.} \red{At the same time, these structures should not be identified one-to-one with finite exotic baryons in vacuum, because the present calculation includes medium effects through infinite matter under periodic boundary conditions and through the effective nonlinearities $\epsilon_i$, while the color and spin correlations entering the EOS are treated in the large-$N$ limit.}

Our model does not exclude the possibility that some of these structures should instead be interpreted as hadronic molecules, i.e., correlations between clusters. Rather, the Y-shaped patterns seem to exhibit a roughly alternating alignment. Nevertheless, since our coherent-state-based formulation does not allow an unambiguous identification of neutron or proton clusters, we leave a detailed study of these correlations for future work.

\begin{figure}[tb]
\includegraphics[width=0.51\textwidth]{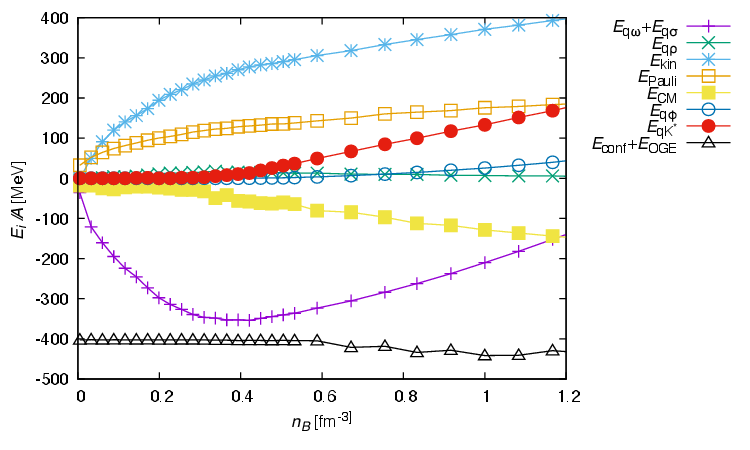}
\caption{\label{fig:Ecomp} Energy contributions as a function of density for the best-performing sampled case, $g_{qK^*}/g_{q\omega}=0.40$: the kinetic energy $E_{\rm kin}$, the confinement energy $E_{\rm conf}$, the one-gluon-exchange energy $E_{\rm OGE}$, the effective Pauli energy $E_{\rm Pauli}$, and the quark-meson exchange energies $E_{\rm qi}$ ($i=\omega,\sigma,\rho,\phi,\text{and }K^*$).}
\end{figure}

At the end of this section, we present in Fig.~\ref{fig:Ecomp} how each energy contribution behaves. As the density increases, the color-magnetic interaction is found to act attractively. Since the figure presents integrated quantities, this cannot be seen unambiguously; however, because the color-magnetic interaction is short-ranged, it plays a critical role in generating multiquark clusters.

\section{DISCUSSION}
In this study, CSMD calculations were used to explore EOS subject to selected astrophysical constraints and bulk nuclear benchmarks.
Three main messages emerge from the present calculation: (i) within this framework, reproducing $M_{\max}>2M_\odot$ requires effective many-body contributions implemented through the nonlinear factors $\epsilon_i$; (ii) the coupling constant $g_{qK^*}$ has a significant impact on stellar radii; and (iii) color-magnetic interactions enhance quark clustering.

For $\epsilon_i=0$, correlations are restricted to pairs of particles only, i.e., there are no effective many-body components. Within the limited framework of the present model, introducing $\epsilon_i$ is therefore essential for reproducing observationally acceptable neutron stars. Similar conclusions about the importance of nonlinear many-body contributions are also found in other EOS models~\cite{DD2, Muto2021, Yamamoto2022}.

The dependence on $g_{qK^*}$ should also be interpreted carefully. In the present work we represent color-singlet channels through an effective quark-meson coupling, but the mapping from baryonic forces to quark-level interactions is intrinsically nontrivial. The sampled value $g_{qK^*}/g_{q\omega}=0.40$ is therefore not a unique best-fit parameter; rather, it is the best-performing case among the discrete sampled values considered here. \red{The present scan does not determine the continuously allowed interval around this value. That preference could shift once additional attractive channels, including a possible attractive $qK$ channel, are included, once both the interaction terms and the kinetic energy are treated on the same relativistic footing, or once future radius measurements more decisively favor either smaller or larger neutron-star radii.}

The clustering result is likewise suggestive rather than definitive. The classification relies on the operational distance criterion $d_{\rm cluster}=0.15$ fm, and the coherent-state formulation does not allow an unambiguous identification of each cluster as a neutron, proton, or specific exotic hadron. \red{Appendix C provides a first threshold-sensitivity check for $d_{\rm cluster}=0.10$ and $0.20$ fm. What appears robust within the present analysis is the trend that color-magnetic interactions enhance correlated color-singlet multiquark structures and suppress isolated quarklike configurations at the highest densities reached on the stable branch, even though the detailed cluster-size distribution remains criterion dependent.} \red{From the perspective of the energy balance, the clearest trend in Fig.~\ref{fig:Ecomp} is that the color-magnetic interaction becomes increasingly attractive with density. We therefore do not emphasize here a directly visible reduction of the integrated Pauli contribution at the onset of strangeness. This interpretation is also consistent with Fig.~\ref{fig:clst_cmag}(c), where clustering still occurs even without strangeness as long as the color-magnetic interaction is retained.} It will be important in future work to clarify how this clustering tendency is related to possible diquark condensation.

It should also be emphasized that the spin discussion in Appendix A provides an effective prescription for estimating pairwise spin-color correlations within the present CSMD framework. A more rigorous implementation of many-body spin correlations remains an important task for future work.

Several extensions are therefore important. Appendix C provides a first sensitivity check with respect to $d_{\rm cluster}$, but a broader threshold scan would further sharpen the phenomenology. It will also be necessary to assess whether the many-body treatment should remain confined to the quark-meson sector or be extended to channels such as $Q+B \to B+Q$, where $Q$ denotes a $q$ or $s$ quark~\cite{Rijken2024}, and to confront the resulting force picture with information on baryonic substructure such as the proton pressure distribution~\cite{Burkert2018,Shanahan2019}. More broadly, the present implementation still has several limitations: the interactions are not yet fully relativistic, finite-temperature and antiparticle effects are omitted, and explicit antisymmetrization remains to be implemented. Future work should address these points, compare with finite-system observables and transport codes such as UrQMD and JAM~\cite{Steinheimer2008, Petersen2008, Akamatsu2018}, and further develop the formalism along the lines discussed in Refs.~\cite{Enyo2005,Enyo2007}. 

\begin{acknowledgments}
We are grateful to T.~Kinugawa, P.~Gubler, T.~Muto, Y.~Yamamoto, T.~Rijken, M.~Oka, and T.~Hatsuda for helpful discussions.
This work was supported by JSPS KAKENHI Grants No. 20H04742, No. 20K03951, and No. 24K07054.
The calculations were performed by the supercomputing system HPE SGI 8600 at the Japan Atomic Energy Agency.
\end{acknowledgments}


\appendix
\section{Effective Spin correlation}

For $N$ quarks with total spin $S$, the sum of spin-correlation factors is given by
\begin{align}
\sum_{i<j} \bm{\sigma}_i \cdot \bm{\sigma}_j
&= \frac{1}{2}\!\left[\left(\bm{\sigma}_1 + \cdots + \bm{\sigma}_N\right)^2 - \sum_{i=1}^N \bm{\sigma}_i^{\,2}\right]  \nonumber \\
&= 2S(S+1) - \frac{3}{2}N. 
\label{eq:gspin}
\end{align}
To calculate the effective spin interaction between up and down, we use the normal Young--Yamanouchi basis. 
The tableau
\begin{center}
\hspace{0em}$\overbrace{\hspace{0.3em}}^{\displaystyle q} \overbrace{\hspace{0.3em}}^{\displaystyle p} $\\
$\begin{array}{|c|c|}
\hline
\cdots & \cdots \\
\hline
\cdots& \multicolumn{1}{c}{} \\
 \cline{1-1}
\end{array}$
\end{center}
has total spin $\frac p 2$. The total number of quarks is $p + 2q$. The first and second rows represent up spins and down spins, respectively.

Using the symmetry of the normal basis, we obtain the effective pairwise spin correlations
\begin{align}
\langle \uparrow\uparrow \rangle = \langle \downarrow\downarrow \rangle
= \frac{1}{4} \langle \bm{\sigma}_\uparrow \cdot \bm{\sigma}_\uparrow \rangle
= \frac{1}{4}, 
\label{eq: spinup}
\end{align}
and
\begin{align}
\langle \uparrow\downarrow \rangle
= \frac{1}{4} \langle \bm{\sigma}_\uparrow \cdot \bm{\sigma}_\downarrow \rangle
= -\frac{1}{2(p+q)} - \frac{1}{4}. 
\label{eq: spindown}
\end{align}

Let us check whether these are consistent with the above general formula (\ref{eq:gspin}).
The number of $\langle \uparrow\uparrow \rangle$ and $\langle \downarrow\downarrow \rangle$ components is
$\binom{p+q}{2} + \binom{q}{2}$, and the number of $\langle \uparrow\downarrow \rangle$ components is $(p + q)q$, so
\begin{align}
\sum_{i<j} \bm{\sigma}_i \cdot \bm{\sigma}_j
&= \frac{1}{2}(p+q)(p+q-1) + \frac{1}{2}q(q-1) \nonumber \\
&~~~~~ + (p+q)q\!\left(-\frac{2}{p+q}-1\right) \nonumber \\
&= 2S(S+1) - \frac{3}{2}N ,
\end{align}
where $N=p+2q,~ S=\tfrac{p}{2}$. Hence these expressions are consistent. 
Therefore $\langle\uparrow\downarrow\rangle \sim -\tfrac{1}{4}$ when the total number of quarks goes to infinity.




The simple effective-pair prescription used above is insufficient for $\Lambda$ without an additional recoupling assumption.
For a spin-$\tfrac{1}{2}$ baryon, there are two Young--Yamanouchi bases;

\ytableausetup{boxsize=1.2em, aligntableaux=center}
\begin{center}
\begin{ytableau}
1 & 2 \\ 3
\end{ytableau},
\qquad
\begin{ytableau}
1 & 3 \\ 2
\end{ytableau}.
\end{center}

The first one is the normal basis. If the wave function is totally antisymmetric, different recoupling bases are equivalent for the total spin operator, although they distribute the pairwise correlations differently. Using the normal basis, we obtain
\begin{align}
\bm{\sigma}_1 \cdot \bm{\sigma}_2 = 1,\qquad
\bm{\sigma}_1 \cdot \bm{\sigma}_3 = \bm{\sigma}_2 \cdot \bm{\sigma}_3 = -2. 
\end{align}

If a strange quark occupies position 3, however, the physical light-quark pair in $\Lambda$ must be in a spin-0 state, so the second basis should be used instead, yielding
\begin{align}
\bm{\sigma}_1 \cdot \bm{\sigma}_2 = -3,\qquad
\bm{\sigma}_1 \cdot \bm{\sigma}_3 = \bm{\sigma}_2 \cdot \bm{\sigma}_3 = 0. 
\end{align}
For the nucleon, one may likewise choose a basis in which a particular $ud$ pair is spin 0, analogous to the $\Lambda$ case. However, because all three quarks are light and degenerate, the color-magnetic mass shift depends only on the total weighted sum of pair correlations, so the recoupling choice does not change the final nucleon mass: both expressions above give the same total $\sum_{i<j}\bm{\sigma}_i\!\cdot\!\bm{\sigma}_j=-3$. For $\Sigma$ baryons, by contrast, the light-quark pair has spin 1, so the normal basis is the appropriate one. The special role of $\Lambda$ is therefore not the existence of an alternative basis itself, but the fact that the strange quark breaks the mass degeneracy. In the $\Lambda$ channel, the hyperfine term must be evaluated with the spin-0 $ud$ pair of the second basis, for which only the $ud$ pair contributes while the $us$ and $ds$ contributions vanish. In this simplified treatment, the hyperfine contribution to $\Lambda$ then matches that of the nucleon, and the remaining mass difference is approximated by the constituent-mass shift, $M_\Lambda - M_N \sim m_s - m$. We use this relation to fix the strange-light mass splitting.

\section{Effective Color correlation}

For $N$ quarks, the sum of color-correlation factors is given by
\begin{align}
\sum_{i<j} \lambda_i^c \lambda_j^c
&= \frac{1}{2}\!\left[\left(\lambda_1 + \cdots + \lambda_N\right)^2 - \sum_{i=1}^N \lambda_i^{2}\right] \nonumber \\
&= 2C_C - \frac{8}{3}N, 
\label{eq:gcolor}
\end{align}
where $C_C$ is the quadratic Casimir of color $\mathrm{SU}(3)$.
We again use the normal Young--Yamanouchi basis for color, with rows for red, green, and blue (with row lengths $p$, $q$, and $r$, respectively, for a total of $p+2q+3r$ quarks), in which the tableau is written as
\begin{center}
\hspace{0em}$\overbrace{\hspace{0.3em}}^{\displaystyle r} \overbrace{\hspace{0.3em}}^{\displaystyle q} \overbrace{\hspace{0.3em}}^{\displaystyle p} $\\
$\begin{array}{|c|c|c|}
\hline
\cdots & \cdots & \cdots \\
\hline
\cdots & \cdots & \multicolumn{1}{c}{} \\
 \cline{1-2}
\cdots& \multicolumn{2}{c}{} .\\
 \cline{1-1}
\end{array}$
\end{center}

As in the spin case, we introduce the effective pair-correlation factors associated with color:
\begin{align}
\langle RR\rangle &= \langle \lambda_R^c \lambda_R^c\rangle = \frac{4}{3} = \langle GG\rangle = \langle BB\rangle, \\
\langle RG\rangle &= -\frac{2}{p+q+r} - \frac{2}{3}, \\
\langle RB\rangle &= -\frac{2}{p+q+r} - \frac{2}{3}, \\
\langle GB\rangle &= -\frac{2}{q+r} - \frac{2}{3}. 
\end{align}

Let us check whether the above result is consistent with the general formula (\ref{eq:gcolor}).
The number of $\langle RR\rangle$, $\langle GG \rangle$, and $\langle BB \rangle$ components is
$\binom{p+q+r}{2} + \binom{q+r}{2} + \binom{r}{2}$,
the number of $\langle RG\rangle$ components is $(p + q + r)(q + r)$, and the number of $\langle GB\rangle$ components is $(q + r)r$.
The effective counting of all pairs then gives
\begin{align}
\sum_{i<j} \lambda_i^c \lambda_j^c
&= \left[ \binom{p+q+r}{2} + \binom{q+r}{2} + \binom{r}{2} \right]\frac{4}{3} \notag \\
&\quad + (p+q+r)(q+r)\!\left(-\frac{2}{p+q+r} - \frac{2}{3}\right) \notag \\
&\quad + (p+q+r)r\!\left(-\frac{2}{p+q+r} - \frac{2}{3}\right) \notag \\
&\quad + (q+r)r\!\left(-\frac{2}{q+r} - \frac{2}{3}\right)  \notag \\
&= 2C_C - \frac{8}{3}N. 
\end{align}
, which is consistent with the general formula (\ref{eq:gcolor}). 
Thus $\langle RR\rangle,\langle GG\rangle,\langle BB\rangle$ converge to $\tfrac{4}{3}$, and
$\langle RG\rangle,\langle RB\rangle,\langle GB\rangle$ converge to $-\tfrac{2}{3}$ in the large-$N$ limit.

For color-singlet states only ($p=0$, $q=0$, $r=\tfrac{N}{3}$), the effective color correlations reduce to
\begin{align}
\langle RR\rangle = \langle GG\rangle = \langle BB\rangle &= \frac{4}{3}, \\
\langle RG\rangle = \langle RB\rangle = \langle GB\rangle &= -\frac{6}{N} - \frac{2}{3}.
\end{align}

\section{\textcolor{red}{Sensitivity to clustering threshold}}

\begin{figure*}[tb]
\begin{flushleft}
(a)-1 \vspace{-1.1cm}
\end{flushleft}
\includegraphics[width=0.96\textwidth]{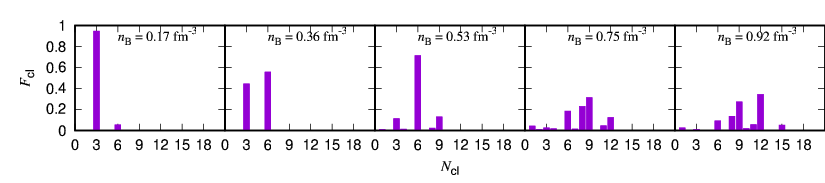} \\
\begin{flushleft}
(a)-2 \vspace{-1.1cm}
\end{flushleft}
\includegraphics[width=0.96\textwidth]{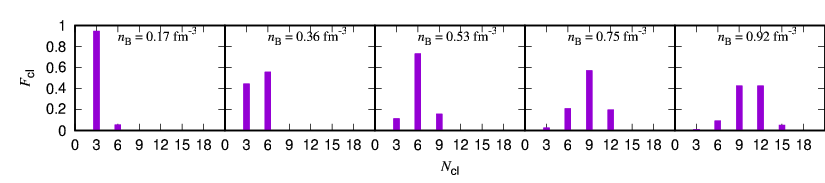} \\
\begin{flushleft}
(b)-1 \vspace{-1.1cm}
\end{flushleft}
\includegraphics[width=0.96\textwidth]{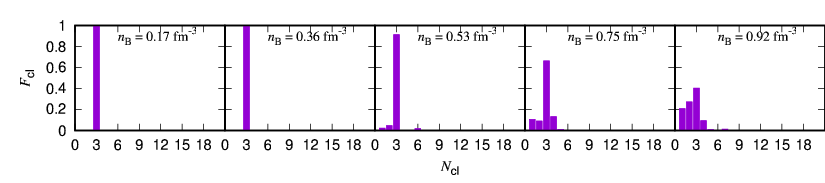} \\
\begin{flushleft}
(b)-2 \vspace{-1.1cm}
\end{flushleft}
\includegraphics[width=0.96\textwidth]{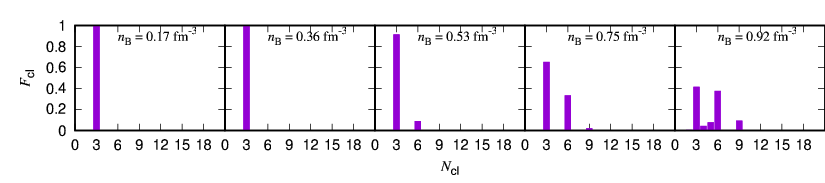}
\caption{\label{fig:clst_thresh}\textcolor{red}{Same as Figs.~\ref{fig:clst_cmag}(a) and (b), but with the clustering threshold changed to $d_{\rm cluster}=0.10$ and $0.20$ fm. Panels (a)-1 and (a)-2 correspond to the case with the color-magnetic interaction for $d_{\rm cluster}=0.10$ and $0.20$ fm, respectively. Panels (b)-1 and (b)-2 correspond to the case without the color-magnetic interaction for $d_{\rm cluster}=0.10$ and $0.20$ fm, respectively. The corresponding result for $d_{\rm cluster}=0.15$ fm is shown in Figs.~\ref{fig:clst_cmag}(a) and (b) in the main text.}}
\end{figure*}

\red{As a simple sensitivity check of the operational clustering criterion, Fig.~\ref{fig:clst_thresh} compares the same configurations as in Figs.~\ref{fig:clst_cmag}(a) and (b), but with $d_{\rm cluster}=0.10$ and $0.20$ fm. These auxiliary choices bracket the adopted value $d_{\rm cluster}=0.15$ fm used in the main text. As expected, the detailed cluster-size distribution changes quantitatively with the threshold. Nevertheless, the qualitative contrast remains visible: with the color-magnetic interaction included, isolated quarklike configurations are suppressed along the stable branch, whereas without it they reappear at the highest density shown. The prominence of $N_{\rm cl}=3n$ is reduced when the clustering threshold is varied away from the adopted value and should therefore be regarded as criterion dependent rather than threshold independent. Accordingly, even though the detailed distributions vary with $d_{\rm cluster}$, the suppression of isolated quarklike configurations by the color-magnetic interaction remains qualitatively visible under the present criterion.}

\newpage

\bibliographystyle{apsrev4-2}
\bibliography{refs}

@article{Riley2019,
  author = {T. E. Riley and others},
  doi = {10.3847/2041-8213/ab481c},
  journal = {Astrophys. J. Lett.},
  volume = {887},
  pages = {L21},
  year = {2019},
}

@article{Miller2019,
  author = {M. C. Miller and others},
  doi = {10.3847/2041-8213/ab50c5},
  journal = {Astrophys. J. Lett.},
  volume = {887},
  pages = {L24},
  year = {2019},
}

@article{Riley2021,
  author = {T. E. Riley and others},
  doi = {10.3847/2041-8213/ac0a81},
  journal = {Astrophys. J. Lett.},
  volume = {918},
  pages = {L27},
  year = {2021},
}

@article{Miller2021,
  author = {M. C. Miller and others},
  doi = {10.3847/2041-8213/ac089b},
  journal = {Astrophys. J. Lett.},
  volume = {918},
  pages = {L28},
  year = {2021},
}

@article{Vinciguerra2024,
  author = {S. Vinciguerra and others},
  doi = {10.3847/1538-4357/acfb83},
  journal = {Astrophys. J.},
  volume = {961},
  pages = {62},
  year = {2024},
}

@article{Choudhury2024,
  author = {D. Choudhury and others},
  doi = {10.3847/2041-8213/ad5a6f},
  journal = {Astrophys. J.},
  volume = {971},
  pages = {L20},
  year = {2024},
}

@article{Abbott2017,
  author = {B. P. Abbott and others},
  doi = {10.1103/PhysRevLett.119.161101},
  journal = {Phys. Rev. Lett.},
  volume = {119},
  pages = {161101},
  year = {2017},
}

@article{Antoniadis2013,
  author = {J. Antoniadis and others},
  doi = {10.1126/science.1233232},
  journal = {Science},
  volume = {340},
  pages = {1233232},
  year = {2013},
}

@article{Nishizaki2002,
  author = {S. Nishizaki and Y. Yamamoto and T. Takatsuka},
  doi = {10.1143/PTP.108.703},
  journal = {Prog. Theor. Phys.},
  volume = {108},
  number = {4},
  pages = {703},
  year = {2002},
}

@article{Vidana2011,
  author = {I. Vida{\~n}a and D. Logoteta and C. Provid{\^e}ncia and A. Polls and I. Bombaci},
  doi = {10.1209/0295-5075/94/11002},
  journal = {Europhys. Lett.},
  volume = {94},
  pages = {11002},
  year = {2011},
}

@article{Yamamoto2014,
  author = {Y. Yamamoto and T. Furumoto and N. Yasutake and Th. A. Rijken},
  doi = {10.1103/PhysRevC.90.045805},
  journal = {Phys. Rev. C},
  volume = {90},
  pages = {045805},
  year = {2014},
}

@article{Lonardoni2015,
  author = {D. Lonardoni and A. Lovato and S. Gandolfi and F. Pederiva},
  doi = {10.1103/PhysRevLett.114.092301},
  journal = {Phys. Rev. Lett.},
  volume = {114},
  pages = {092301},
  year = {2015},
}

@article{Togashi2016,
  author = {H. Togashi and E. Hiyama and Y. Yamamoto and M. Takano},
  doi = {10.1103/PhysRevC.93.035808},
  journal = {Phys. Rev. C},
  volume = {93},
  pages = {035808},
  year = {2016},
}

@article{Bednarek2012,
  author = {I. Bednarek and P. Haensel and J. L. Zdunik and M. Bejger and R. Ma{\'n}ka},
  doi = {10.1051/0004-6361/201118560},
  journal = {A\&A},
  volume = {543},
  pages = {A157},
  year = {2012},
}

@article{Weissenborn2012,
  author = {S. Weissenborn and D. Chatterjee and J. Schaffner-Bielich},
  doi = {10.1103/PhysRevC.85.065802},
  journal = {Phys. Rev. C},
  volume = {85},
  pages = {065802},
  year = {2012},
}

@article{Muto2022,
  author = {T. Muto and T. Maruyama and T. Tatsumi},
  doi = {10.1093/ptep/ptac115},
  journal = {Prog. Theor. Exp. Phys.},
  pages = {093D03},
  year = {2022},
}

@article{Ma2023,
  author = {F. Ma and C. Wu and W. Guo},
  doi = {10.1103/PhysRevC.107.045804},
  journal = {Phys. Rev. C},
  volume = {107},
  pages = {045804},
  year = {2023},
}

@article{Huang2025,
  author = {C. Huang and L. Tolos and C. Provid{\^e}ncia and A. Watts},
  doi = {10.1093/mnras/stae2792},
  journal = {Mon. Not. R. Astron. Soc.},
  volume = {536},
  pages = {3262},
  year = {2025},
}

@article{Burgio2021,
  author = {G. F. Burgio and H.-J. Schulze and I. Vida{\~n}a and J.-B. Wei},
  doi = {10.1016/j.ppnp.2021.103879},
  journal = {Prog. Part. Nucl. Phys.},
  volume = {120},
  pages = {103879},
  year = {2021},
}

@article{Vidana2025,
  author = {I. Vida{\~n}a and V. Mantovani Sarti and J. Haidenbauer and D. L. Mihaylov and L. Fabbietti},
  doi = {10.1140/epja/s10050-025-01539-z},
  journal = {Eur. Phys. J. A},
  volume = {61},
  pages = {59},
  year = {2025},
}

@article{Akimura2005,
  author = {Y. Akimura and T. Maruyama and N. Yoshinaga and S. Chiba},
  doi = {10.1140/epja/i2005-10143-x},
  journal = {Eur. Phys. J. A},
  volume = {25},
  pages = {405},
  year = {2005},
}

@article{Yasutake2024,
  author = {N. Yasutake and T. Maruyama},
  doi = {10.1103/PhysRevD.109.043056},
  journal = {Phys. Rev. D},
  volume = {109},
  pages = {043056},
  year = {2024},
}

@article{Wanajo2024,
  author = {S. Wanajo and S. Fujibayashi and K. Hayashi and K. Kiuchi and Y. Sekiguchi and M. Shibata},
  doi = {10.1103/PhysRevLett.133.241201},
  journal = {Phys. Rev. Lett.},
  volume = {133},
  pages = {241201},
  year = {2024},
}

@article{DD2,
  author = {S. Typel and G. R{\"o}pke and T. Kl{\"a}hn and D. Blaschke and H. H. Wolter},
  doi = {10.1103/PhysRevC.81.015803},
  journal = {Phys. Rev. C},
  volume = {81},
  pages = {015803},
  year = {2010},
}

@article{Chatziioannou2025,
  author = {K. Chatziioannou and others},
  doi = {10.1103/RevModPhys.97.045007},
  journal = {Rev. Mod. Phys.},
  volume = {97},
  pages = {045007},
  year = {2025},
}

@article{Yamamoto2022,
  author = {Y. Yamamoto and N. Yasutake and Th. A. Rijken},
  doi = {10.1103/PhysRevC.105.015804},
  journal = {Phys. Rev. C},
  volume = {105},
  pages = {015804},
  year = {2022},
}

@article{Nanamura2022,
  author = {T. Nanamura and others},
  doi = {10.1093/ptep/ptac101},
  journal = {Prog. Theor. Exp. Phys.},
  pages = {093D01},
  year = {2022},
}

@article{MaruyamaHatsuda2000,
  author = {T. Maruyama and T. Hatsuda},
  doi = {10.1103/PhysRevC.61.062201},
  journal = {Phys. Rev. C},
  volume = {61},
  pages = {062201},
  year = {2000},
}

@article{YasutakeMaruyama2024,
  author = {N. Yasutake and T. Maruyama},
  doi = {10.11804/NuclPhysRev.41.QCS2023.06},
  journal = {Nucl. Phys. Rev.},
  volume = {41},
  pages = {801},
  year = {2024},
}

@article{MaruyamaYasutake2024,
  author = {T. Maruyama and N. Yasutake},
  doi = {10.11804/NuclPhysRev.41.QCS2023.07},
  journal = {Nucl. Phys. Rev.},
  volume = {41},
  pages = {806},
  year = {2024},
}

@article{Aaron2020,
  author = {A. Park and S.-H. Lee and T. Inoue and T. Hatsuda},
  doi = {10.1140/epja/s10050-020-00078-z},
  journal = {Eur. Phys. J. A},
  volume = {56},
  pages = {93},
  year = {2020},
}

@article{Wu2025,
  author = {C. Wu and W. Guo},
  doi = {10.1140/epjp/s13360-025-06145-y},
  journal = {Eur. Phys. J. Plus},
  volume = {140},
  pages = {193},
  year = {2025},
}

@article{Stone2007,
  author = {J. Rikovska Stone and P. A. M. Guichon and H. H. Matevosyan and A. W. Thomas},
  doi = {10.1016/j.nuclphysa.2007.05.011},
  journal = {Nucl. Phys. A},
  volume = {792},
  pages = {341},
  year = {2007},
}

@article{Vinas2014,
  author = {X. Vi{\~n}as and M. Centelles and X. Roca-Maza and M. Warda},
  doi = {10.1140/epja/i2014-14027-8},
  journal = {Eur. Phys. J. A},
  volume = {50},
  pages = {27},
  year = {2014},
}

@article{Li2019,
  author = {B.-A. Li and P. G. Krastev and D.-H. Wen and N.-B. Zhang},
  doi = {10.1140/epja/i2019-12780-8},
  journal = {Eur. Phys. J. A},
  volume = {55},
  pages = {117},
  year = {2019},
}

@article{Hornick2018,
  author = {N. Hornick and L. Tolos and A. Zacchi and J.-E. Christian and J. Schaffner-Bielich},
  doi = {10.1103/PhysRevC.98.065804},
  journal = {Phys. Rev. C},
  volume = {98},
  pages = {065804},
  year = {2018},
}

@article{Danielewicz2017,
  author = {P. Danielewicz and P. Singh and J. Lee},
  doi = {10.1016/j.nuclphysa.2016.11.008},
  journal = {Nucl. Phys. A},
  volume = {958},
  pages = {147},
  year = {2017},
}

@article{Estee2021,
  author = {J. Est{\'e}e and others},
  doi = {10.1103/PhysRevLett.126.162701},
  journal = {Phys. Rev. Lett.},
  volume = {126},
  pages = {162701},
  year = {2021},
}

@article{Baym1971,
  author = {G. Baym and C. Pethick and P. Sutherland},
  doi = {10.1086/151216},
  journal = {Astrophys. J.},
  volume = {170},
  pages = {299},
  year = {1971},
}

@article{Caplan2018,
  author = {M. E. Caplan and A. S. Schneider and C. J. Horowitz},
  doi = {10.1103/PhysRevLett.121.132701},
  journal = {Phys. Rev. Lett.},
  volume = {121},
  pages = {132701},
  year = {2018},
}

@article{Bauswein2017,
  author = {A. Bauswein and O. Just and H.-T. Janka and N. Stergioulas},
  doi = {10.3847/2041-8213/aa9994},
  journal = {Astrophys. J. Lett.},
  volume = {850},
  pages = {L34},
  year = {2017},
}

@article{Shibata2019,
  author = {M. Shibata and E. Zhou and K. Kiuchi and S. Fujibayashi},
  doi = {10.1103/PhysRevD.100.023015},
  journal = {Phys. Rev. D},
  volume = {100},
  pages = {023015},
  year = {2019},
}

@article{Shahrbaf2022,
  author = {M. Shahrbaf and D. Blaschke and S. Typel and G. R. Farrar and D. E. Alvarez-Castillo},
  doi = {10.1103/PhysRevD.105.103005},
  journal = {Phys. Rev. D},
  volume = {105},
  pages = {103005},
  year = {2022},
}

@article{Xia2025,
  author = {C. Xia and X. Lai and R. Xu},
  doi = {10.1142/S0217751X25501805},
  journal = {Int. J. Mod. Phys. A},
  volume = {40},
  number = {34},
  pages = {2550180},
  year = {2025},
}

@article{Muto2021,
  author = {T. Muto and T. Maruyama and T. Tatsumi},
  doi = {10.1016/j.physletb.2021.136587},
  journal = {Phys. Lett. B},
  volume = {820},
  pages = {136587},
  year = {2021},
}

@unpublished{Rijken2024,
  author = {Th. A. Rijken},
  note = {Report THEF-NIJM 24.03, unpublished},
  url = {https://nn-online.org/eprints/pdf/24.03.pdf},
  year = {2024},
}

@article{Burkert2018,
  author = {V. D. Burkert and L. Elouadrhiri and F. X. Girod},
  doi = {10.1038/s41586-018-0060-z},
  journal = {Nature},
  volume = {557},
  pages = {396},
  year = {2018},
}

@article{Shanahan2019,
  author = {P. E. Shanahan and W. Detmold},
  doi = {10.1103/PhysRevLett.122.072003},
  journal = {Phys. Rev. Lett.},
  volume = {122},
  pages = {072003},
  year = {2019},
}

@article{Steinheimer2008,
  author = {J. Steinheimer and M. Bleicher and H. Petersen and S. Schramm and H. St{\"o}cker and D. Zschiesche},
  doi = {10.1103/PhysRevC.77.034901},
  journal = {Phys. Rev. C},
  volume = {77},
  pages = {034901},
  year = {2008},
}

@article{Petersen2008,
  author = {H. Petersen and J. Steinheimer and G. Burau and M. Bleicher and H. St{\"o}cker},
  doi = {10.1103/PhysRevC.78.044901},
  journal = {Phys. Rev. C},
  volume = {78},
  pages = {044901},
  year = {2008},
}

@article{Akamatsu2018,
  author = {Y. Akamatsu and M. Asakawa and T. Hirano and M. Kitazawa and K. Morita and K. Murase and Y. Nara and C. Nonaka and A. Ohnishi},
  doi = {10.1103/PhysRevC.98.024909},
  journal = {Phys. Rev. C},
  volume = {98},
  pages = {024909},
  year = {2018},
}

@article{Enyo2005,
  author = {Y. Kanada-En'yo and O. Morimatsu and T. Nishikawa},
  doi = {10.1103/PhysRevC.71.045202},
  journal = {Phys. Rev. C},
  volume = {71},
  pages = {045202},
  year = {2005},
}

@article{Enyo2007,
  author = {Y. Kanada-En'yo and O. Morimatsu and T. Nishikawa},
  doi = {10.1143/PTPS.168.194},
  journal = {Prog. Theor. Phys. Suppl.},
  volume = {168},
  pages = {194},
  year = {2007},
}
\end{document}